\documentclass[11pt, oneside]{article}

\usepackage[dvipsnames,usenames]{color}
\usepackage{graphicx}				
\DeclareGraphicsExtensions{.pdf, .jpg, .png}
\graphicspath{{Figures/}}

\usepackage{amssymb,amsmath,amsthm}

\title{Household Income Distribution in the USA}
\author{Costas J. Efthimiou and Adam Wearne\\
Department of Physics\\ University of Central Florida\\ Orlando, FL 32816}

\date{December 20, 2015}							

\begin{document}
\maketitle

\begin{quote}
In this article we present an alternative model for the distribution of household incomes in the United States.  We provide arguments from two differing perspectives which both yield the proposed income distribution curve, and then fit this curve to empirical data on household income distribution obtained from the United States Census Bureau.  
\end{quote}

\section{Introduction}

Understanding the statistical nature of \textit{income} and \textit{wealth} distributions has been a long-standing problem in the field of economics.  Income is based on the concept of \textit{money}.  Although related to each other, the two concepts---income and wealth---are not  interchangeable.  In this article, we are interested in the study of  income distributions in the USA.

One of the earliest and most notable attempts at understanding  wealth---the separation from income was fuzzier then---was pioneered by Pareto 
\cite{Pareto}. Pareto realized that the density $\rho(w)$ of people per unit wealth $w$, when $w$ is a relative large number, follows a simple power law
$$
       \rho(w)  \propto {1\over w^{\alpha+1}} .
$$  
From his data, Pareto estimated the exponent $\alpha$ to be about $3/2$. Today the above relation is known as \textit{Pareto's law}
and the exponent $\alpha$ as the \textit{Pareto exponent}. 
Since the time of Pareto, many improvements and alternative approaches have  been proposed 
as well as extensive discussions on income and wealth inequality. (Among a huge list, perhaps references \cite{AB1, AB2, 
CCC, Mitzenmacher} is a good starting point.)  

Gibrat \cite{Gibrat} was actually  the first to look at the income of the middle-range people and conclude that that the distribution follows a log-normal curve.  
The Pareto and Gibrat conclusions have been derived mathematically  and studied more carefully by many mathematicians and
physicists (for example, \cite{Champ,Gabaix,Kesten,Simon,Sornette}).  They have been at the core of the most of the traditional models of income which treat individual income as a stochastic variable and from which the income distribution is derived based on an analysis of
 the behavior at low and high income values and, eventually, interpolating between between the two ranges. 
 It is now known that the Kesten stochastic process, defined by the recursion $x_{i+1}= a_i \, x_i + b_i$,  where $a_i$, $b_i$ are 
 positive random  numbers  can produce a power law in the tail.  The role of the $b_i$  term is of paramount importance;  it was  quickly understood to be necessary in order to generate the power law. Without it, the distribution of $x_i$ is log-normal.

More recently, with the explosion of econophysics, empirical studies of income distributions have been carried out but many physicists
who have used ideas from physics to analyze and interpret the data. Among them are  those by Dragulescu and Yakovenko 
\cite{Dragulescu2000,Dragulescu2001} for UK and USA, 
by Fujiwara et al. \cite{FSAKA} for Japan, 
 by Nirei and Souma  \cite{NS} for US and Japan,
 by Ferrero \cite{Ferrero1, Ferrero2} for Japan, UK, New Zealand and Argentina. 
Also, Clementi and collaborators have looked at the power tails in income distributions for Italy, Germany, UK and USA \cite{CG, CG2, CMG}.

The knowledge of a population's distribution of income is an important piece of information to understand  the economic health of a nation through an objective  quantitative tool.  It also provides one with the means of estimating income inequality, and can reveal information on the existence of economic classes within a society.  Developing a theoretical model of society to predict the form of income distributions is then an important step in accurately determining the aforementioned data which can then be used as a guide to help build a secure and more stable economic landscape.

Excluding the high income range, Dragulescu and Yakovenko have argued that there is evidence for an  exponential Boltzmann distribution of income for single earner households  \cite{Dragulescu2000}.  The idea and data are reviewed in  \cite{Dragulescu2003, Yakovenko2009}. 
Motivated by the belief that economic markets and systems that are comprised by a large number of interacting economic
agents must be describable by the the same statistical laws obeyed by physical systems composed of interacting particles, physicists have created many kinetic models of wealth and income.  In such models, agents interact via an exchange process wherein they exchange some amount of their wealth in a process which is analogous to the exchange of momentum between gas particles. A sample of papers of such models are \cite{BY, BT,
Chakraborti2000, CCM, CC, CGK, Cordier2005, Patriarca2010};
the books \cite{CCCC,PT} summarize very nicely the kinetic exchange models created by physicists to describe income and wealth
 distributions (where additional related references may also be found).
Inspired by this trend, in the present article, we propose  a model for the income distribution of households in the United States.  Our model is developed based on analogy with physical systems
--- in particular,  the blackbody radiation \cite{Mandle} --- and provides an excellent fit to US Census data of income distributions.  This approach of transferring ideas from physical systems to economic systems (when and where possible) grants us the advantage of accessing the large body of mathematical techniques developed in statistical physics while simultaneously providing a clear interpretation of the parameters involved in our system.

\section{The Model of Income  Distribution}
\label{sec:TheModel}

Consider all possible income states of an economic agent. It is immediate that they  constitute a discrete and, in principle, infinite set.
Let $s_1,  s_2,  s_3, \dots $ be these states and $r_1 < r_2 <r_3 <\dots$ respectively be  the agent's income in these states.
We also  assume that the values $r_1, r_2, \dots$  are time-independent.  
 
Now imagine an economic society made of $N$ agents. This society is described by its own states of income $S_1, S_2, S_3, \dots$ 
which can be related to the states of a single agent as follows.  Let $n_i$ agents having income $r_i$, $i=1,2,\dots$. We will call the $n_i$ the 
\textit{occupation number}. Obviously,  
$$
    N = \sum_i n_i .
$$
The total income of the society is
$$
   R = \sum_i  n_i \, r_i ,
$$
Even for an isolated society, the numbers $N$ and $R$ do not have to be fixed. Besides human births and deaths, a developed economic society allows the creation of corporations which can have income. However, the numbers $N$ and $R$ are not crucial. The states
$S_\alpha$ of the society are characterized by the collection of the occupation numbers $n_1, n_2, n_3, \dots$.
 
\subsection*{Method 1}
Consider the states $s_m$ and $s_p$  of a single economic agent with income values $r_m < r_p$. Let their occupation numbers for the society be $n_m$ and $n_p$ respectively.
It is possible that some agents in state $s_p$  lose some of their income without  \textit{any particular financial reason} and drop in state 
$s_m$. For example farmers may see their crops destroyed by
weather conditions and hence they will not be able to sell them to get income.  `Weather conditions' have nothing to do with established
financial activities, so we consider them `no particular financial reason'.  Such activities which create losses without financial reasons,  we call  \textbf{spontaneous drops}.  The probability for a spontaneous drop to happen, during an infinitesimal time 
interval $dt$, is
$$
    dP(p\to m)= A_m^p \, dt ,
$$
where $A_m^p$ is some coefficient. 

On the other hand, we can have agents who  transition from one state to  another because of financial activities. For example, an employer can hire a number of agents to work for him. When he pays the employees, his income decreases and the income of each of the employees increases.  
Such  income changes which happen under the action of a financial  activity, we call \textbf{stimulated drops} if the income decreased  or \textbf{stimulated raises} if the income increased. The probability for a stimulated drop to happen, during an infinitesimal time  interval $dt$, is
$$
    dP(p\to m) = B_m^p \, \rho_p \, dt ,
$$
where $B_m^p$ is some coefficient,
$$
      \rho(r) = {dn \over dr}.
$$
 is the  \textbf{income distribution density} --- density of the number of agents per income and $\rho_p$ is a shorthand
 notation for $\rho(r_p)$. The determination of the density $\rho(r)$ is the focus of the current paper. 

Similarly, for a stimulated raise, we have
$$
    dP(p\leftarrow m)= B^m_p \, \rho_m \, dt ,
$$
where $B_p^m$ is yet another coefficient.
Notice that, in the above discussion, we have excluded spontaneous raises. No agent can enjoy an increase in income  without a 
particular financial reason.

From the above discussion, we conclude that the total probability for an income drop $p\to m$ is
$$
      dP_\text{total}(p\to m)= ( A_m^p + B_m^p \, \rho_p)  \, dt .
$$
The corresponding population change is
$$
      dn(p\to m)=  n_p ( A_m^p + B_m^p \, \rho_p)  \, dt .
$$
For the raise, the corresponding population change is
$$
      dn(p\leftarrow m)=  n_m \, B^m_p \, \rho_m  \, dt .
$$
For a society in equilibrium,  the occupation numbers remain more or less fixed. Hence
$$
      n_p ( A_m^p + B_m^p \, \rho_p)  \, dt =       n_m \, B^m_p \, \rho_m  \, dt .
$$
At the same time, for large populations  the occupation numbers follow the Boltzmann distribution\cite{Mandle,Yakovenko2000,Yakovenko2009}:
\begin{eqnarray*}
   n_p  = {Ng_p\over Z} \, e^{-\beta r_p}, \quad
   n_m = {Ng_m\over Z} \, e^{-\beta r_m} ,
\end{eqnarray*}
where $\beta$ is the inverse temperature (a measure of the average income), $Z$ a normalization constant (known as partition function),
$$
     Z =  \sum_p  g_p \, e^{-\beta r_p},
$$
 and  $g_p, g_m$  constants which take into account possible degeneracies in the states of a single agent.
So, finally the equilibrium condition is:
$$
      g_p \, e^{-\beta r_p} ( A_m^p + B_m^p \, \rho_p)   =       g_m \, e^{-\beta r_m}\, B^m_p \, \rho_m   .
$$
This relation is true for any  income density $\rho$. In particular, a society in which the average income is infinite: $\rho\to+\infty$ and  $\beta\to 0$. Hence,
$$
   g_p B_m^p = g_m B^m_p .
$$
Now, let's take two successive states $p=m+1$. Hence $r_{m+1}-r_m=r$ and $\rho_{m+1}=\rho_m=\rho(r)$.  Then
$$
      \rho(r) = {C(r) \over e^{\beta r} - 1} ,
$$
where $C(r)=A_m^{m+1} / B_m^{m+1}$.

The ratio $A_m^p / B_m^p$ for any income states $p$ and $m$ should be computed through a model  based on socio-economic ideas and actual data. We have constructed no such model. However,  it seems natural to assume that the ratio in the
income density has a simple power dependence, that is   
 $$
      C(r) = c \, r^\alpha , 
$$
 where $\alpha$ and $c$ some constants. We have thus concluded that the income density  of a society should be given by a function of the form
\begin{equation}
      \rho(r) = {c \, r^\alpha \over e^{\beta r} - 1} .
\label{eq:density1}
\end{equation}

\subsection*{Method 2}

The partition function for the society is
$$
     Z_\text{society} =  \sum_\alpha g_\alpha \, e^{-\beta R_\alpha} ~.
$$
Any state of income of ths society is characterized by the occupation numbers $n_1, n_2, \dots$ each of which takes all values
0, 1, 2, \dots Hence
\begin{eqnarray*}
     Z_\text{society}(\beta) =  \sum_\alpha g_\alpha \, e^{-\beta R_\alpha} 
                              &=&  \sum_{n_1=0}^\infty  e^{-\beta r_1n_1}  \,  \sum_{n_2=0}^\infty  e^{-\beta r_2n_2} \dots \\
                              &=&  \prod_{i=1}^\infty {1\over 1- e^{-\beta r_i}}  .
\end{eqnarray*}
The average occupation number $n_1$ is 
\begin{eqnarray*}
     \overline{n}_1 &=&  \sum_{n_1=0}^\infty  n_1\,e^{-\beta r_1n_1}  \,  \sum_{n_2=0}^\infty  e^{-\beta r_2n_2} \cdots \\
                            &=&  -{1\over\beta}{\partial\over\partial r_1}\sum_{n_1=0}^\infty  e^{-\beta r_1n_1}  \,  \sum_{n_2=0}^\infty  e^{-\beta r_2n_2} \dots \\
                            &=&  {1\over e^{\beta r_1}-1}  ,
\end{eqnarray*}
and similarly for any other $\overline{n}_i$.

The density $\rho(r)$ is then equal to 
$$
      \rho(r) = \overline{n}(r)  \, g(r) =  {g(r)\over e^{\beta r}-1} ,
$$
where $g(r)$ is the degeneracy at value $r$. Again, this function must be constructed as a result of a socio-economic model. 
This can be a very complicated task and here we assume, similarly to the previous approach, that it is a simple power function:
$$
       g(r) =  c \, r^\alpha ,
$$
with $c$ and $\alpha$ constants to be computed. We thus arrive at the same density function \eqref{eq:density1}.

\subsection*{Actual Density for USA}

Comparison with real data points out that $\alpha=3/2$ (see Section \ref{sec:Validation}). Hence,
\begin{equation}
      \rho(r) = {c \, r^{3/2} \over e^{\beta r} - 1} .
\label{eq:density2}
\end{equation}
Obviously the number of economic agents is the integral of the density over all possible values of income:
$$
     N =  \int_0^{+\infty} \rho(r) \, dr 
        =  {c \over \beta^{5/2}} \, \int_0^{+\infty}  {x^{3/2} \over e^{x} - 1} dx .
$$
Also, the  income $R$ of the society is 
$$
     R =  \int_0^{+\infty}  r \, \rho(r) \, dr 
        =  {c \over \beta^{7/2}} \, \int_0^{+\infty}  {x^{5/2} \over e^{x} - 1} dx .
$$
Let
$$
     I(\alpha) =  \int_0^{+\infty}  {x^\alpha \over e^{x} - 1} dx .
$$
This integral is known to be related to the $\zeta$-function:
$$
     I(\alpha) = \Gamma(\alpha+1) \, \zeta(\alpha+1) ~.
$$
Hence in our case,
$$
     N =   {3c\sqrt{\pi} \over 4\beta^{5/2}} \,  \zeta(5/2) ~ ,
$$
and
$$
     R =   {15c\sqrt{\pi} \over 8\beta^{7/2}} \,  \zeta(7/2) ~ ,
$$
since $\Gamma(5/2)=3\sqrt{\pi}/4$ and $\Gamma(7/2)=(5/2) \cdot \Gamma(5/2)$ .
Writing the last results in the form
$$
     {1\over\beta}  =  \text{const.} \,   N^{2/5} ~,
$$
and
$$
     R  =   \text{const.} \, N\,  {1\over\beta}  ~,
$$
we see that the average income per individual scales with the population as a power law while
the income of the society is proportional to its population.

\section{Validation of the Model}
\label{sec:Validation}

\begin{figure}[p]
\begin{center}
\includegraphics[width=6.2cm]{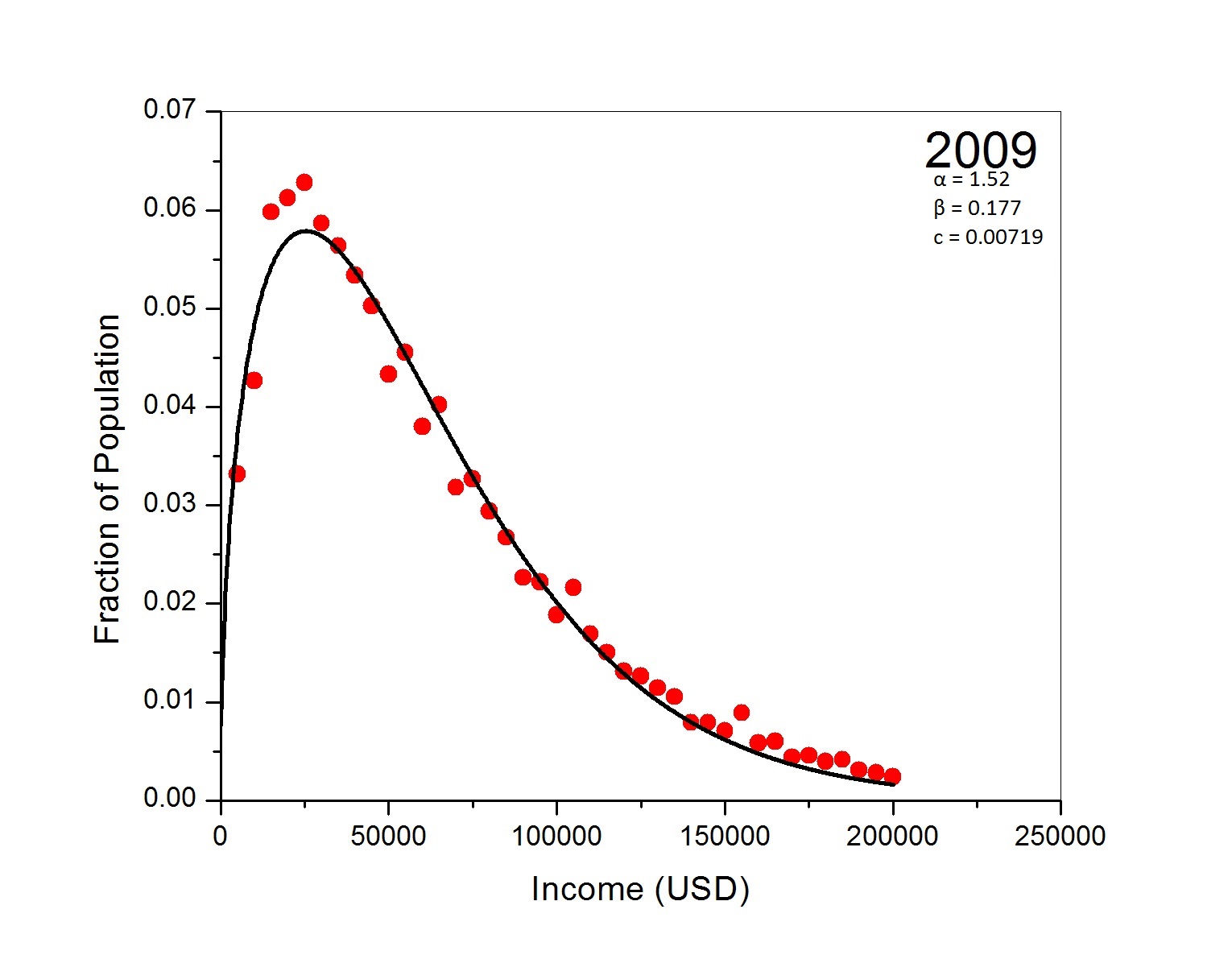}
\includegraphics[width=6.2cm]{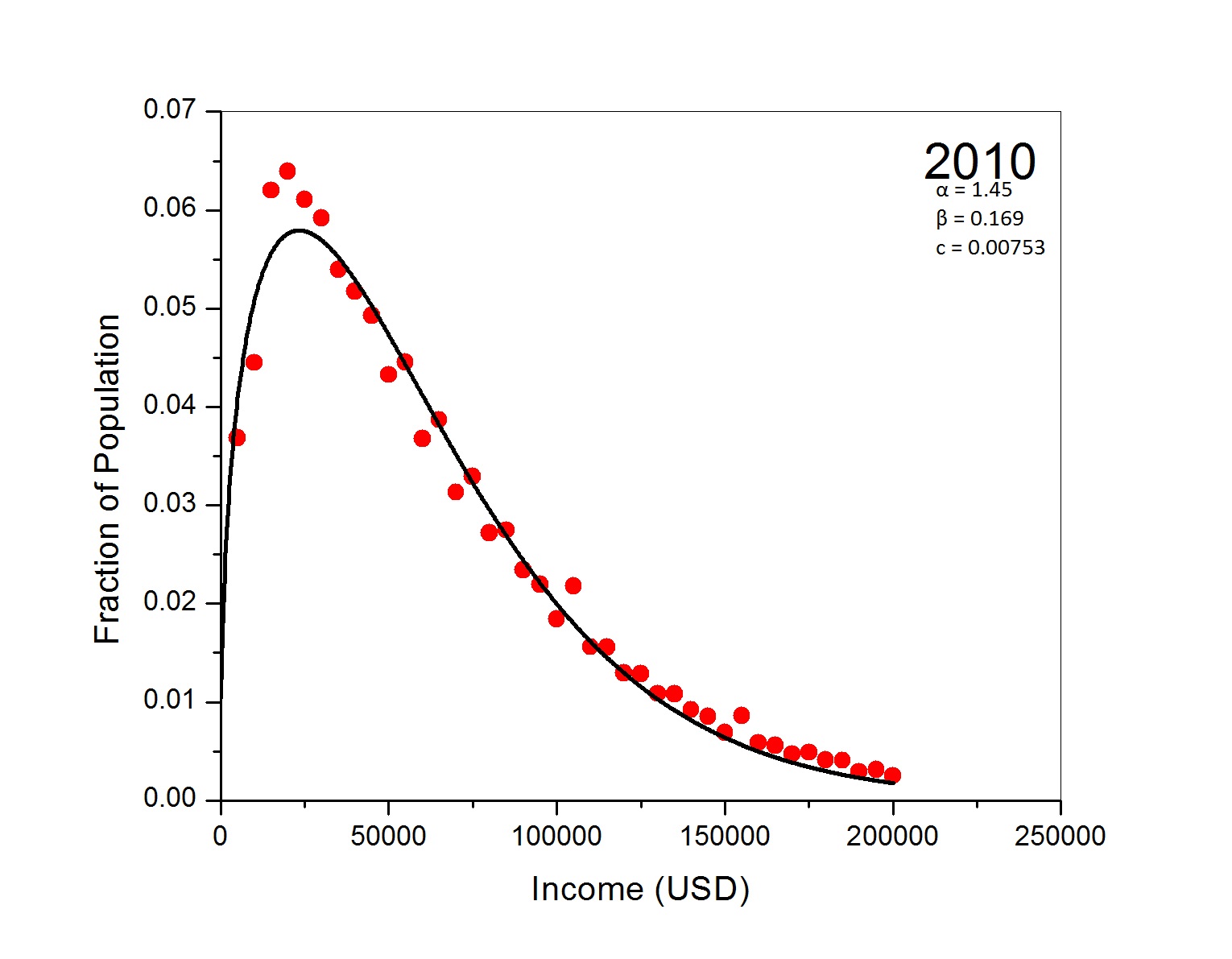}\\
\includegraphics[width=6.2cm]{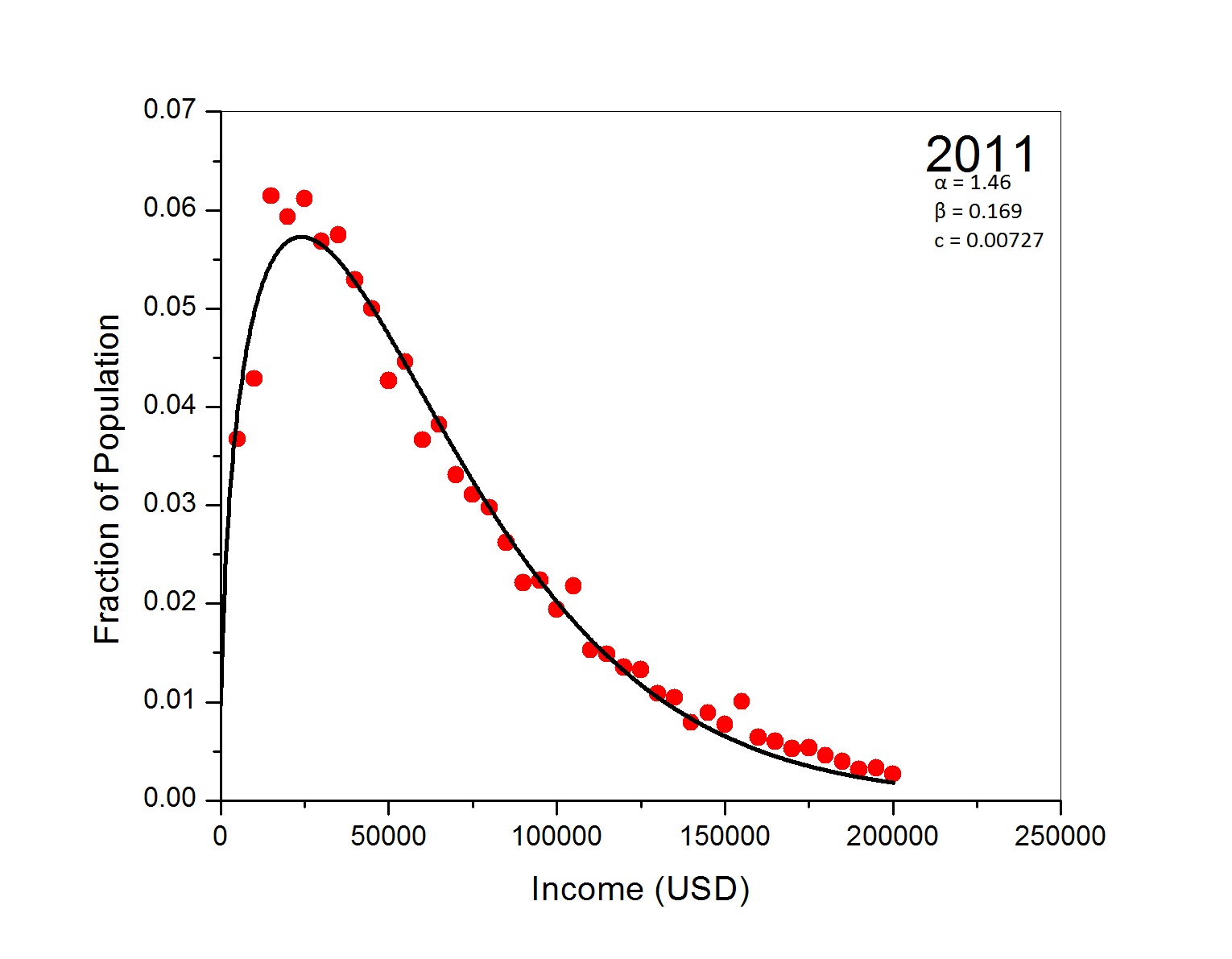}
\includegraphics[width=6.2cm]{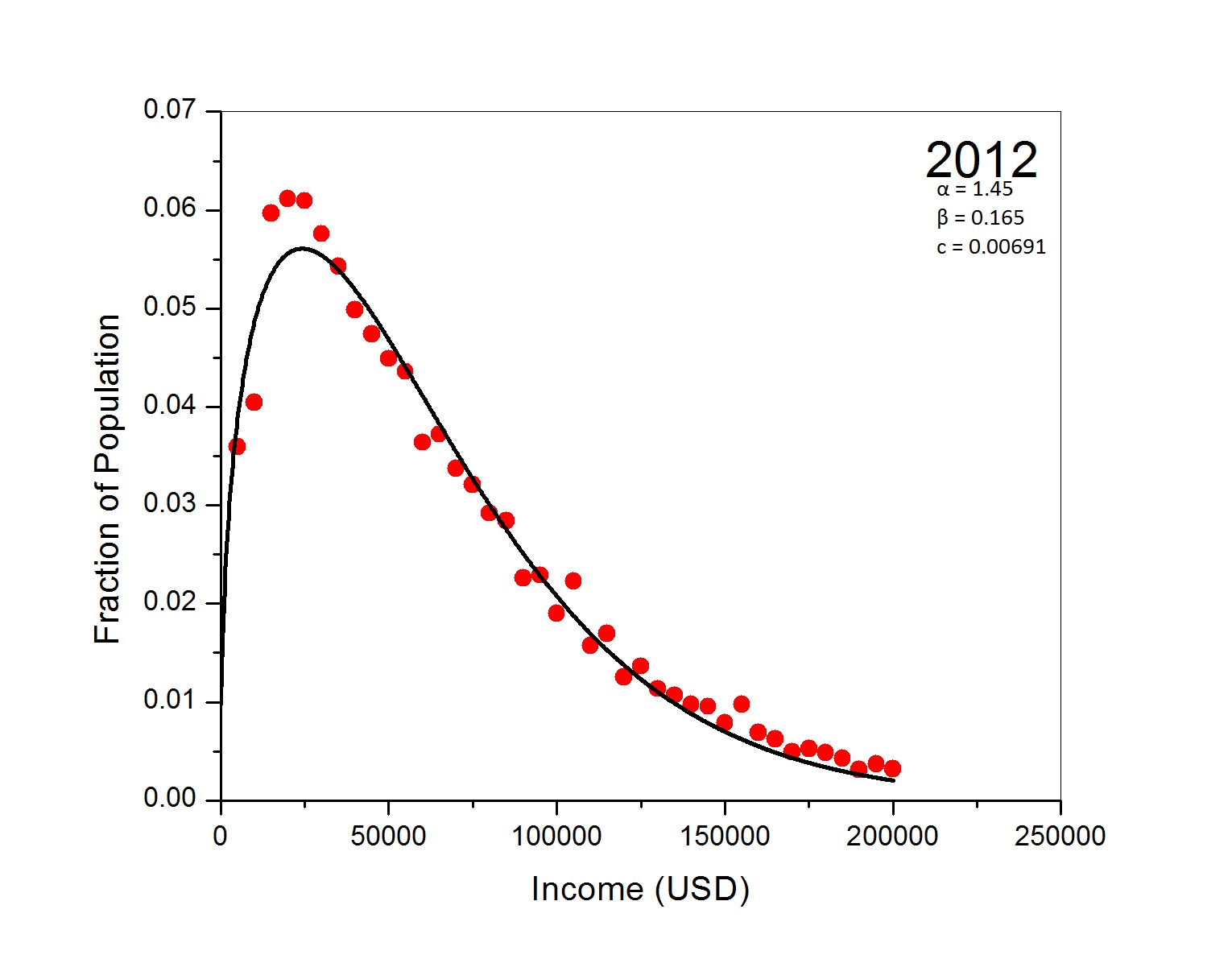}\\
\includegraphics[width=6.2cm]{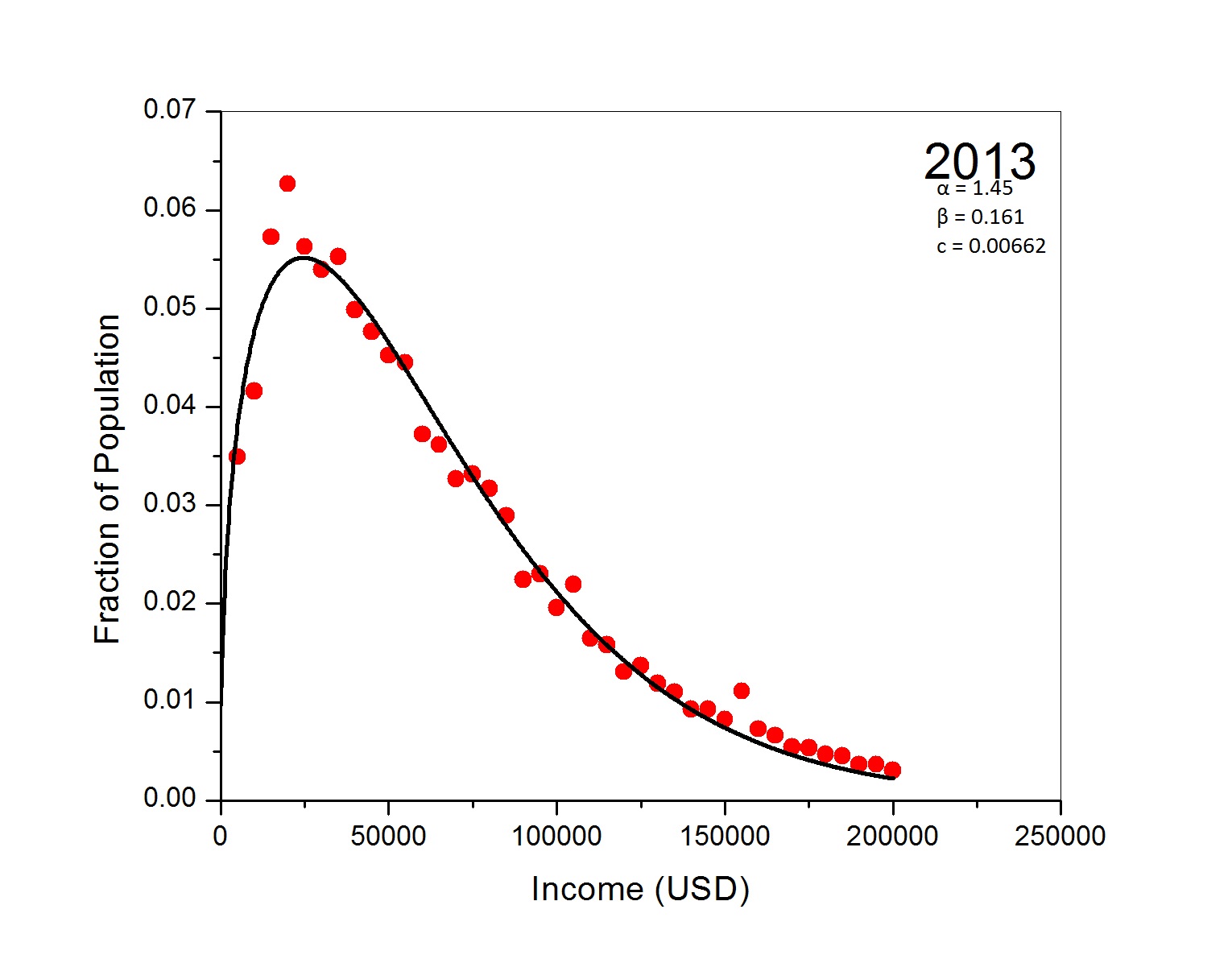}
\end{center}
\caption{\footnotesize Plot of the density function $\rho(r)$ for the years 2009--2013.}
\label{fig:Densities}
\end{figure}

The data used to validate the proposed model --- equation  \eqref{eq:density1} ---  were household income  data \cite{USincome}  obtained from the US Census Bureau \cite{USCensus}.  Incomes ranged up to \$100,000 for years 1994 -- 2008 and up to 
\$200,000 for years 2009 -- 2013.  For each year, the total number of reported household incomes was rescaled to unity for ease of comparison across years.  In other words,
$$
       \int_0^{+\infty} \rho \, dr = 1 .
$$
In the graphs of Figure \ref{fig:Densities}, this amounts to divide the population numbers for each income bracket by the total number of households.
To avoid overpopulating the article with many similar graphs, we have given the income density function $\rho(r)$ for five years,
from 2009 through 2013.  
The data agree with the curve given by equation \eqref{eq:density1} with the earliest and latest years giving the best fits. It is worthwhile to note that the latest years have an impressive agreement with the theoretical model. In  the years omitted, the data appear more scattered  as we move closer to the recession period. Perhaps this behavior of income can be used 
to predict when we move towards a recession with a lead time of a few years. However, it is not known to us how US Census made the
set of data, so some caution should be exercised for its interpretation.

\begin{figure}[h!]
\begin{center}
\setlength{\unitlength}{1mm}
    \begin{picture}(120,40)
        \put(0,-10){ \includegraphics[width=6cm]{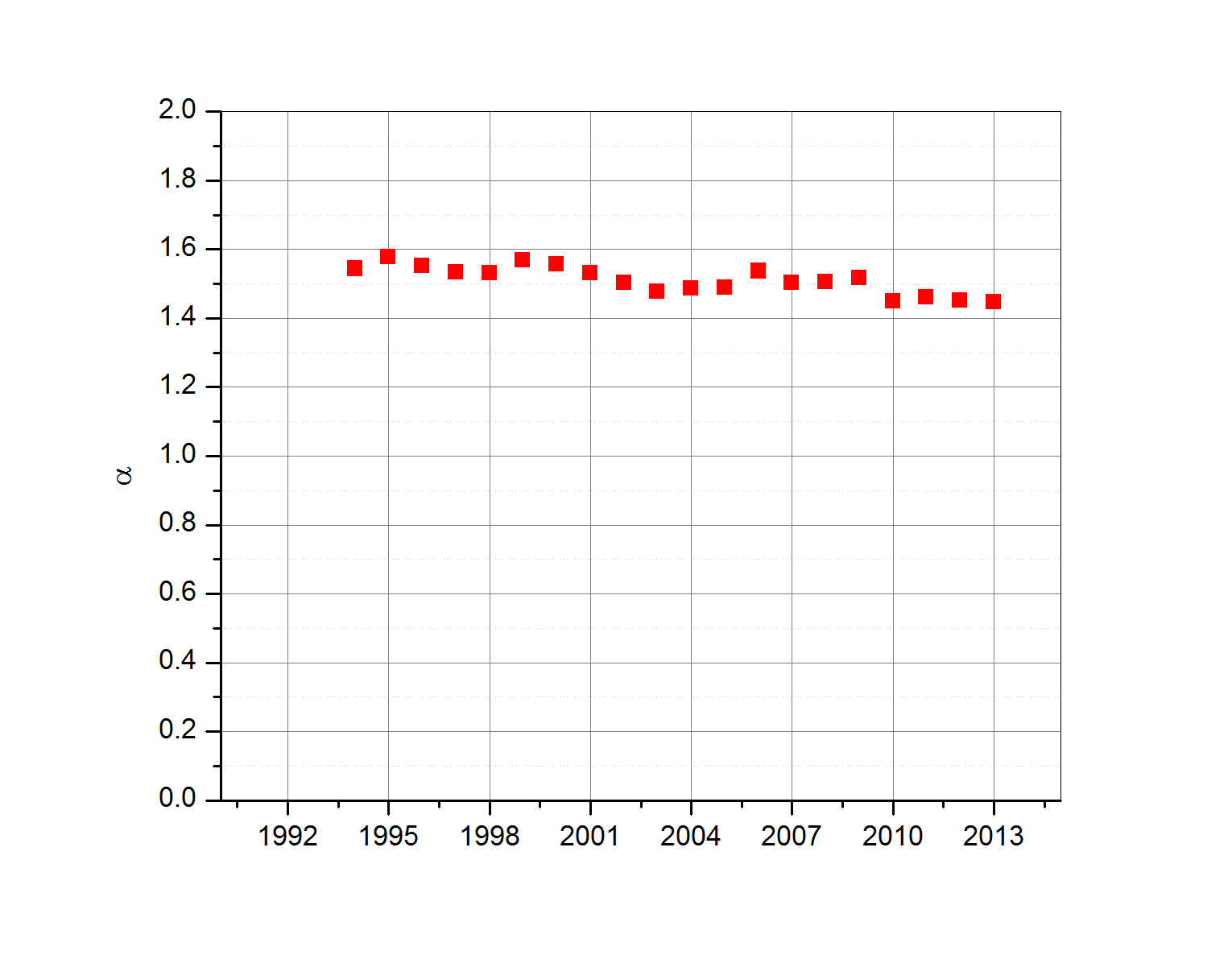}}
        \put(60,-10){\includegraphics[width=6cm]{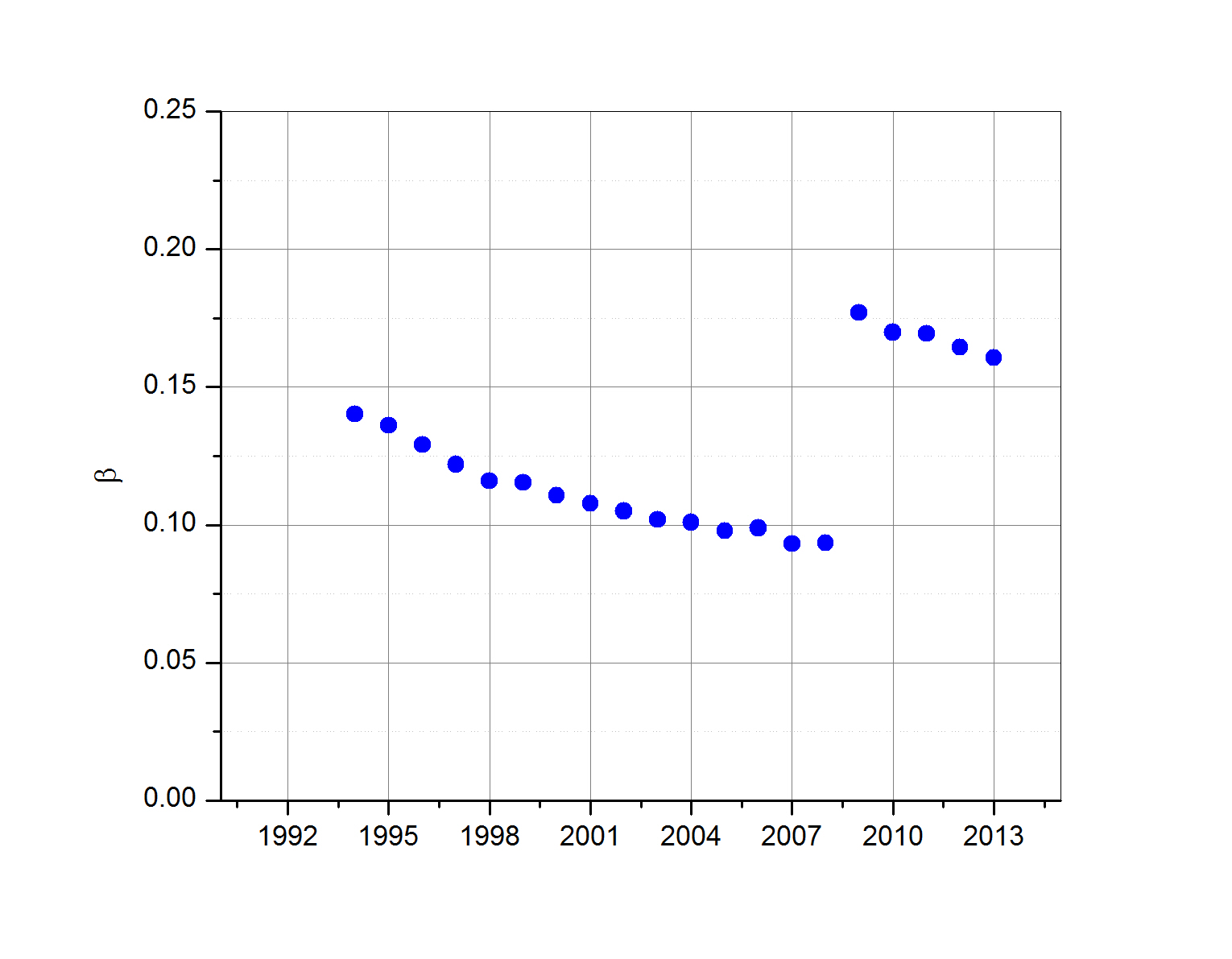}}
    \end{picture}
\end{center}
\caption{\footnotesize Plots of the constants $\alpha$ and $\beta$  for the period 1994--2013.}
\label{fig:AandB}
\end{figure}

\begin{figure}[h!]
\begin{center}
\setlength{\unitlength}{1mm}
    \begin{picture}(120,40)
        \put(0,-10){ \includegraphics[width=6cm]{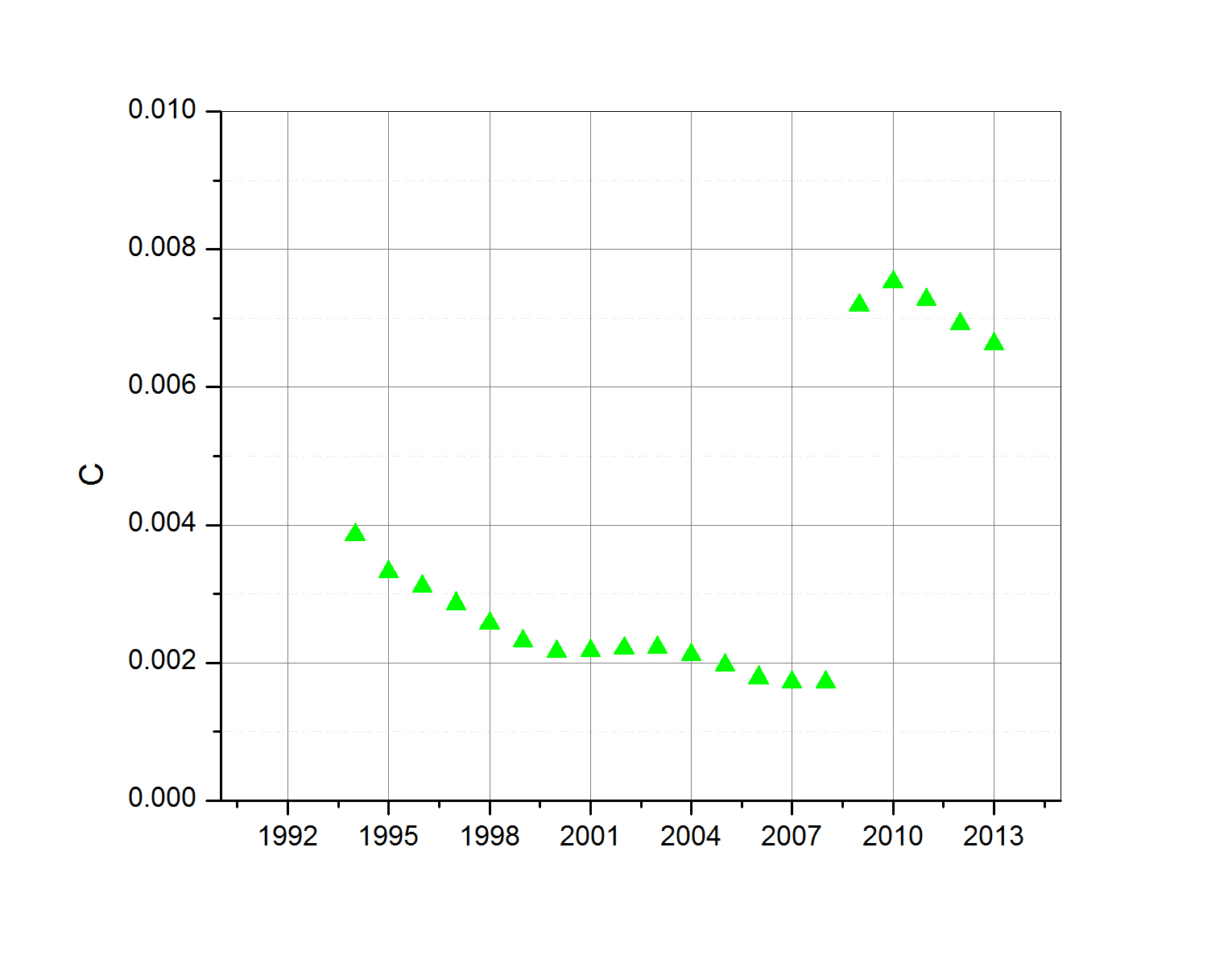}}
        \put(60,-10){\includegraphics[width=6cm]{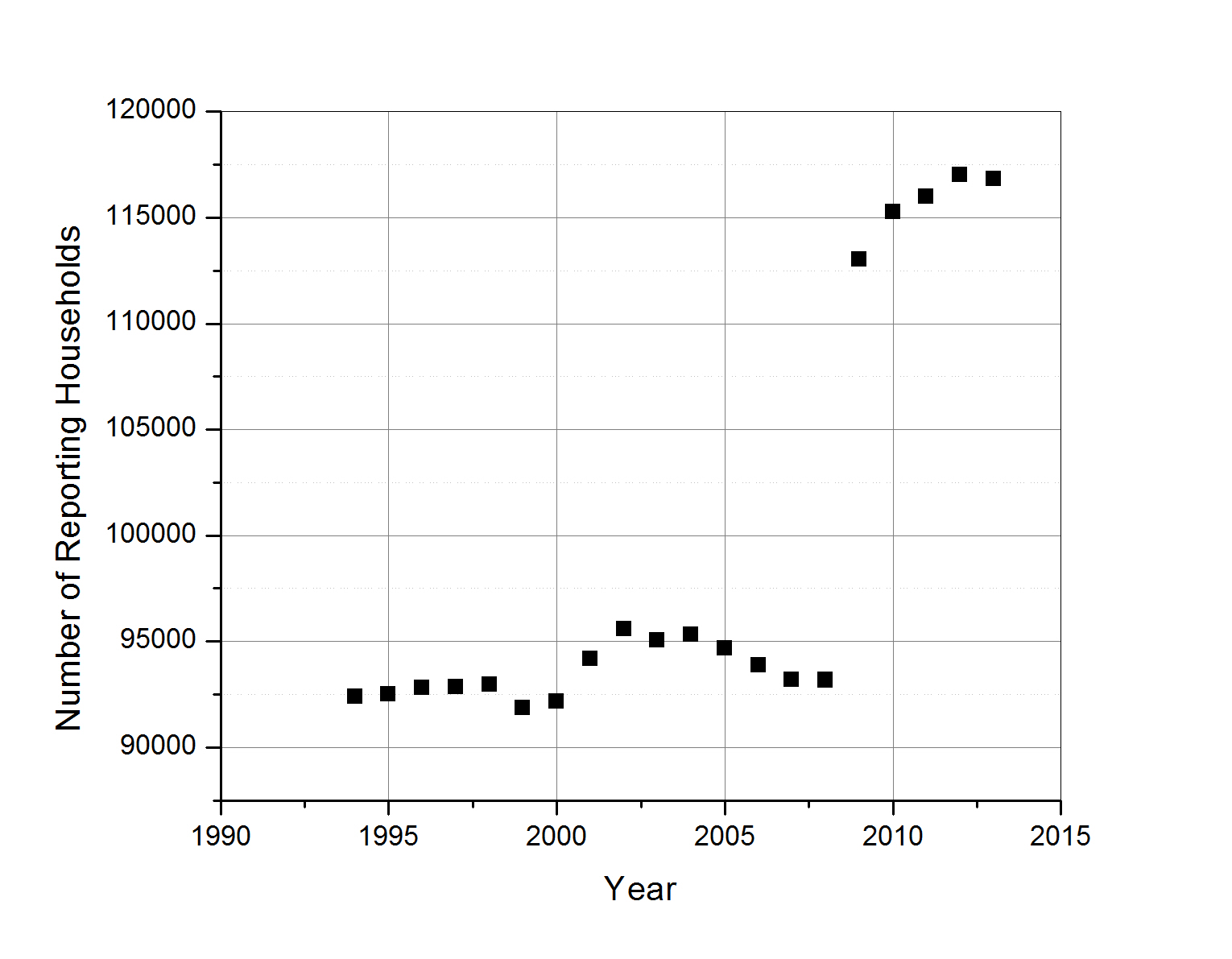}}
    \end{picture}
\end{center}
\caption{\footnotesize Plots of the constant  $c$ and the population $N$ of the economic agents for the period 1994--2013.}
\label{fig:Population}
\end{figure}

From the fitting of the actual data, we can extract the values of the constants $c, \beta, \alpha$. In Figure \ref{fig:AandB}
we give the values of $\alpha$ and $\beta$  over the period 1994 -- 2013 and in  Figure \ref{fig:Population}  the values of $c$ 
and $N$ over the same period.
As a result of the fitting, some very interesting and surprising features  emerge and are worth noting. The exponent $\alpha$, although it fluctuates slightly, appears to 
do so around the value 3/2.  For the parameters $N$, $\beta$ and $c$, there is an abrupt change (discontinuity) at the year 2009 which is the year of financial recovery. Notice that the number of economic agents jumps at higher values; however 
 the average income (represented by $1/\beta$) jumps at lower values. The recession (left side of the discontinuity) however progressed smoothly with no discontinuity appearing at any particular year. 

Finally, we would like to mention that the proposed model has a coefficient of determination $R^2$ that is closer to 1 (perfect fit) than the popular
household model  given by the gamma distribution $c\,r\,e^{-\beta r}$ (See, for example,  \cite{Dragulescu2000}.)  The values of $R^2$
for our model and the gamma model are shown in Table \ref{table:R2}.
\begin{table}[h!]
\begin{center}
\begin{tabular}{|c|c|c|} \hline 
Year  &     $R^2$ (our model)  & $R^2$ (gamma model) \\ \hline
1994	 &	0.983036	 & 0.977200 \\
1995	 &	0.986101	 & 0.981773 \\
1996	 &	0.981993	 & 0.976437 \\
1997	 &	0.982164	 & 0.975105 \\
1998	 &	0.981842	 & 0.974415 \\
1999	 &	0.983001	 & 0.978122 \\
2000	 &	0.980909	 & 0.975126 \\
2001	 &	0.979007	 & 0.971966 \\
2002	 &	0.975130	 & 0.966285 \\
2003	 &	0.976875	 & 0.966319 \\
2004	 &	0.977173	 & 0.967276 \\
2005	 &	0.970794	& 0.960924 \\
2006	 &	0.965006	& 0.958164 \\
2007 &	0.961827	& 0.953202 \\
2008	 &	0.962009	& 0.953932 \\
2009	 &	0.994079	& 0.987708 \\
2010	 &	0.993583	& 0.982211 \\
2011 &	0.993908	& 0.983525 \\
2012	 &	0.992951	& 0.981809 \\
2013	 &	0.993912	& 0.981772 \\ \hline
\end{tabular}
\end{center}
\caption{\footnotesize The  coefficient of determination $R^2$  for our model and a popular household model  given by the gamma distribution $c\,r\,e^{-\beta r}$.  Our model consistently gives values closer to 1 for all years.}
\label{table:R2}
\end{table}
%

\section{Discussion and Conclusions}

In this article we have presented two arguments for obtaining a model of income distributions based on arguments similar to those for obtaining the black-body curve.  For $e^{\beta r}\gg 1$, our model produces a gamma-like distribution curve 
$$
       \rho(r) = c \, r^\alpha \, e^{-\beta r}  ,
$$
which is favored by various authors (for example, see \cite{Angle, CCCC, Ferrero1, Ferrero2}).  We have  shown how the empirical data of a certain range of income values agrees well with the proposed model.   The variation of the model parameters over the years reflects the period of economic instability that the United States experienced between 2003 and 2008.  Interestingly, despite this unstable period, the exponent $\alpha$ of the model remains approximately constant over the years under consideration, equal to 3/2.  
The constancy of this parameter may then provide some deeper insight to the underlying behavior behind income dynamics of households in developed countries. It would be nice if a socio-economic model based on the particulars of the operation of the American society is found to explain the degeneracy $g(r)\propto r^{3/2}$ of income states.  A finite discontinuity in the remaining two parameters of the model, $c$ and $\beta$, appears to locate the exact year of economic recovery, which in our case is 2009.
During deteriorating economic conditions, the data points of the income distribution become more scattered although the profile
of the  curve remains the same.
It will be quite valuable to compare our  data and findings with similar data and findings 
of household income distributions from similarly economically developed countries, as well as from countries which are less developed and face larger economic turbulence.

Finally, we would like to point out the following fact: It is true that in economics and finance, there is a large number of parameters which
appear to allow many different models to be constructed. However, relative good agreement with data is not automatic unless the underlying
assumptions capture some of the characteristics of the dynamics. We thus hope that our model does contain some elements of truth which
time will eventually verify.

\section*{Acknowledgements}

We would like to thank Professor Tristan H\"ubsch,  Dr. Dan Pirjol and the anonymous referees for comments and corrections on the article, 
as well as providing some references which were unknown to us. \\[2mm]

\noindent Both authors contributed equally to this paper.


\end{document}